\documentstyle[twocolumn,prl,aps,epsfig]{revtex} 

\newcommand{\be}{\begin{equation}}
\newcommand{\en}{\end{equation}}
\newcommand{\bea}{\begin{eqnarray}}
\newcommand{\ena}{\end{eqnarray}}

\begin{document}

\draft

\title{Presence of Many Stable Nonhomogeneous States in 
an Inertial Car-Following Model}   

\author{Elad Tomer$^1$, Leonid Safonov$^{1,2}$ and Shlomo Havlin$^1$}
\address{$^1$ Minerva Center and Department of Physics,
         Bar--Ilan~University,
         52900~Ramat--Gan, Israel}
\address{$^2$ Department of Applied Mathematics and Mechanics, Voronezh State
University, 394693~Voronezh, Russia}

\date{\today}

\maketitle

\begin{abstract}

A new single lane car following model of traffic flow is presented. The model 
is inertial and free of collisions. It demonstrates experimentally observed 
features of traffic flow such as the existence of three regimes: free,
fluctuative (synchronized) and congested (jammed) flow; bistability of free
and fluctuative states in a certain range of densities, which causes the
hysteresis in transitions between these states; jumps in the density-flux
plane  in the fluctuative regime and gradual spatial transition from
synchronized to free flow. Our model suggests that in the fluctuative
regime there exist  many stable states with different wavelengths, and that
the velocity fluctuations in the congested flow regime decay approximately
according to a power law in time.

\end{abstract}



In the last years, growing effort has been made in understanding 
traffic flow dynamics. Recent experiments \cite{KR1,KR2} show that
traffic flow demonstrates complex physical phenomena, among which 
are: 
\begin{itemize}
\item[(i)] The existence of three states: free flow, "synchronized" (or
"fluctuative") flow and traffic jams (for low, intermediate and high
densities correspondingly). 
The second state has two essential features: synchronization of flow in
different lanes (for the multilane traffic) and fluctuation performed by the 
system in density-flux plane. Since our model is single-lane we will refer to
this state as "fluctuative".  
\item[(ii)] Hysteresis which is observed in transitions between the free and
the  fluctuative  flow.
\item[(iii)] Long survival time of traffic jams.  
\end{itemize}
Modeling of traffic flow is traditionally performed using two approaches.
The microscopic, or car-following models approach, which describes the nearest-neighbor interaction  
between two consecutive cars and investigates its influence on the 
flow (see e.g. [3-5]), and the macroscopic, or
continuous models approach, which represents the flowing traffic as a 
continuous media and describes it using the hydrodynamical partial
differential equations (see e.g. [7-9]). Wide surveys of
these models are given in [10-12]. 

In this Letter we inroduce an inertial single lane car-following model, which
is free of collisions.  We study the model both numerically and
analytically and find the existence of three regimes in traffic flow:
free flow regime at low densities (where each car moves with almost a constant
velocity), fluctuative flow regime at inetrmediate densities (where stable
periodic oscillations of velocities of all cars are observed) and congested or
jammed flow regime at high densities (where due to high density all the cars
tend to move with the same, relatively small velocity). Our model predicts the
existence of many inhomogeneous stable states in the fluctuative regime and
demonstrates hysteresis in transitions between free and fluctuative 
regimes. The experimentally observed long survival time of jams may be
explained by our finding that the fluctuations in the congested flow regime
decay slowly according to a power law.   

To formulate the model we assume that car acceleration is affected by three
factors: \begin{itemize}
\item[(a)] aspiration to keep safety time gap from the car ahead,
\item[(b)] pre-braking if the car ahead is much slower,
\item[(c)] aspiration not to exceed significantly the permitted velocity.
\end{itemize}
In mathematical description, the acceleration of the $n$th car $a_n$ is 
given by a sum of three terms depending on its coordinate $x_n$, velocity
$v_n$, distance to the car ahead $\Delta x_n=x_{n+1}-x_n$ and the
velocities difference $\Delta v_n=v_{n+1}-v_n$ :
\be
\label{eqModel}
 a_n =  A(1-{{\Delta x_n^0}\over{\Delta x_n}} ) - 
{ {Z^2(-\Delta v_n)} \over {2(\Delta x_n - D)} } -
kZ(v_n-v_{per}),
\en
where $A$ is a sensitivity parameter, $D$ is the minimal 
distance between consecutive cars, $v_{per}$ is the permitted 
velocity and $k$ is a constant. The safety distance 
$\Delta x_n^0=v_nT+D$ depends on $T$, which is the safety time gap
constant. The function $Z$ is defined as $Z(x)=(x+|x|)/2$.
Note that Eq.(\ref{eqModel}) can be generalized by adding a noise term.

In the following we discuss in more details the terms in
the right side of (\ref{eqModel}): 

\begin{itemize}
\item[(a)] The first term plays an important role when
velocity difference between consequtive cars  is relatively small. 
In this case the $n$th car accelerates if $\Delta x_n > \Delta x_n^0$
and brakes if $\Delta x_n < \Delta x_n^0$.

The choice of function in this term is
not unique.  Replacing it by other functions of $\Delta x_n$ which
are increasing, equal to zero if $\Delta x_n=\Delta x_n^0$ and tend to
$-\infty$ if $\Delta x_n\to 0$, such as  $A\log({ \Delta x_n / \Delta
x_n^0})$, gives similar results.  

\item[(b)] The second term plays an important role when $v_n \gg v_{n+1}$. 
A car getting close to a much slower car starts braking
even if $\Delta x_n > \Delta x_n^0$. This term corresponds to the negative 
acceleration needed to reduce $|\Delta v_n|$ to $0$ as $\Delta x_n\to D$.

\item[(c)] The dissipative third term is a repulsive force
acting when the velocity exceeds the permitted velocity.
\end{itemize}

Unlike optimal velocity models \cite{Sug} 
the acceleration in our model depends explicitly
on $\Delta x$ which enables us to make the flow free of collisions.

The motion of cars is therefore described by the following system of ordinary
differential equations

\be
  \label{Initial}
  \left\{
    \begin{array}{lcl}
       \dot x_n & = & v_n, \\
        &&\\
       \dot v_n & = & A(1-\frac{v_nT+D}{x_{n+1}-x_n})\\
        &&\\
        && \quad-\frac{Z^2(v_n-v_{n+1})}{2(x_{n+1}-x_n-D)}-kZ(v_n-v_{per}),
    \end{array}
  \right.
\en 
$\quad n=1,2,\ldots N$ with periodic boundary conditions
$$
\quad x_{N+1} = x_1 +\frac{N}{\rho},\quad v_{N+1}=v_1.
$$

A solution of Eqs. (\ref{Initial}) which corresponds to homogeneous
flow is

\be
\label{Hom_Sol}
v_n^0 = v^0 =
\left\{
\begin{array}{l}
\frac{A(1-D\rho)+kv_{per}}{A\rho T+k},\quad \rho\leq\frac{1}{D+Tv_{per}},\cr
\cr
\frac{1-D\rho}{\rho T},\quad \rho\geq\frac{1}{D+Tv_{per}},
\end{array}
\right.
\en

$$
x_n^0=\frac{n-1}{\rho}+v^0t.
$$

In the following numerical results we use parameters values 
$v_{per}=25(m/s)$, $T=2(s)$, $D=5(m)$, 
$1\leq A\leq 5 (m/s^2)$ and $k=2(s^{-1})$.
 
The flux-density relation (often called the fundamental diagram) for the
homogeneous flow is shown in Fig.1(a) as a dashed line. Comparison of this
curve with the fundamental diagrams (solid lines) obtained by the numerical
solution of equations (\ref{Initial}) for different values of $A$ starting
from nonhomogeneous initial conditions indicates that for values of $\rho$
smaller than some critical value $\rho_1$ or greater than another critical
value $\rho_2$ the flux is the same, while for the intermediate values of
density ($\rho_1<\rho<\rho_2$) the measured flux is considerably lower than the
homogeneous solution flux. Plotting the variance of velocities 
$\sigma_v=[\frac{1}{N}\sum\limits_{n=1}^N(v_n- \langle v\rangle)^2]^{1/2}$
(where $\langle v\rangle$ is the average velocity) against $\rho$ (Fig.1(b))
shows the existence of velocity fluctuations for $\rho_1<\rho<\rho_2$. We can
therefore define three regimes in traffic flow:  the free flow regime
($\rho<\rho_1$), the fluctuative flow regime ($\rho_1<\rho<\rho_2$) and the
congested flow regime ($\rho>\rho_2$). Note that the flow in the first and the
last regimes is homogeneous. Note also that for small values of $A$ $\rho_2$
is greater than the maximal possible density $\rho_{max}=1/D$ and the
congested flow regime does not exist. See Fig. 1(b) for $A=2$. This finding is
supported by the analytical results shown below.

\begin{figure*}
\label{1}

\centerline{
\epsfig{file=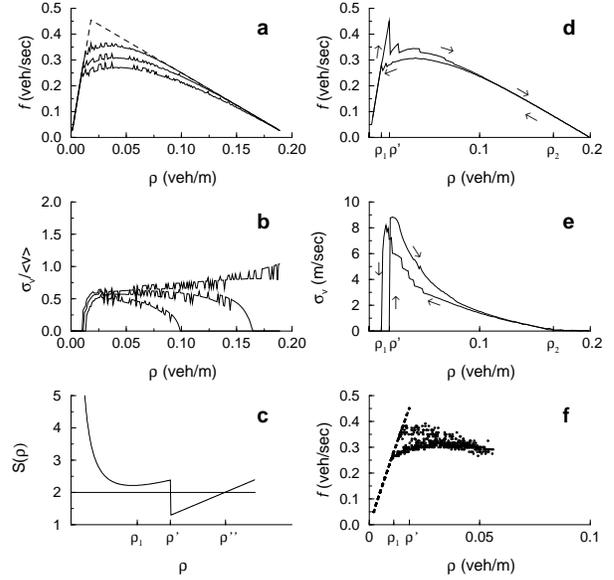, bb= 30 80 600 560, width=8.5cm, angle=270}
}
\vskip 0.3cm
\caption {(a) Fundamental diagram for $A=5,3,2(m/s^2)$ (top to bottom). Dashed
line corresponds to the homogeneous solution. (b) Ratio of variance of
velocities to the average velocity for $A=2,3,5(m/s^2)$ (top to bottom). (c)
Qualitative plot of function $S(\rho)$. (d),(e) Hysteresis loops in
transitions between free and fluctuative flow states for $A=3$ , arrows show
the direction of changing the global density. (f) Results of local
measurements of density and flux in free (almost straight line) and
fluctuative regimes.}

\end{figure*}

In order to estimate the values of $\rho_1$ and $\rho_2$ we analyse the
stability of the homogeneous flow  solution.
The linearization of Eqs. (\ref{Initial}) 
near the homogeneuos flow solution (\ref{Hom_Sol}) 
in variables $\xi_n = x_n - x_n^0$ has the form

\be
  \label{GenLin}
  \ddot\xi_n=-p\dot\xi_n+q(\xi_{n+1}-\xi_n),\quad n=1,\ldots,N,
\en
$$
\xi_{N+1}=\xi_1,
$$

where $p=AT\rho+k$, $q=\frac{AT+kTv_{per}+kD}{AT\rho+k}\cdot A\rho^2$
 for $\rho\leq\frac{1}{D+Tv_{per}}$ and 
$p=AT\rho$, $q=A\rho$ otherwise.

A solution of equation (\ref{GenLin}) can be written as

\be
 \label{Solution}
  \xi_n=\exp\{i\alpha n+ zt\},
\en
where $\alpha=\frac{2\pi}{N}\kappa$ ($\kappa=0,\ldots,N-1$) and $z$ - a
complex number. Substituting (\ref{Solution}) into (\ref{GenLin}) we obtain the
algebraic equation for $z$ 

\be
 \label{CharEq}
  z^2+pz-q(e^{i\alpha}-1)=0.
\en

Each of the $N$ equations (\ref{CharEq}) has two solutions. These $2N$
different complex numbers are the eigenvalues of system (\ref{GenLin}).
One of them (which corresponds to $\kappa=0$) is equal to zero regardless of
values of parameters. In this case all $\xi_n$ in (\ref{Solution})
are equal to a constant and belong to the
one-dimensional subspace of equilibria of system (\ref{GenLin}) (defined by
equations $\xi_1=\ldots=\xi_N$, $\dot\xi_1=\ldots=\dot\xi_N=0$).
This indicates that the disturbed state $x_n$ for $z=0$ is also
homogeneous. For $z\neq 0$ $\xi_n$ in (\ref{Solution}) is a wave with
increasing or decreasing amplitude. Therefore, if we find conditions under
which other $2N-1$ eigenvalues have negative real parts (the magnitude of wave
(\ref{Solution}) decreases with time) we can say that under these conditions
the homogeneous flow solution (\ref{Hom_Sol}) is stable. 

Following the approach of \cite{Sug} we can derive this condition as
$\frac{p^2}{q}>2$ or $S(\rho)>2$, where

\be
S(\rho)=
\left\{
\begin{array}{l}
\frac{(AT\rho+k)^3}{\rho^2A(AT+kv_{per}T+kD)},\quad\rho\leq\frac{1}{D+Tv_{per}},\cr
  \cr
 A\rho T^2,\quad
\rho\geq\frac{1}{D+Tv_{per}}. \end{array}
\right.
\en

A qualitative plot
of $S(\rho)$ is sketched in Fig.1(c). From this figure it follows that
depending on $\rho$ we have three regimes of
stability/instability of the homogeneous flow solution. If $\rho<\rho'$ (free
flow) or $\rho>\rho''$ (congested flow) the homogeneous flow solution is stable
and if $\rho'<\rho<\rho''$ it is unstable, where $\rho'=\frac{1}{D+Tv_{per}}$
and $\rho''=\frac{2}{AT^2}$. Note that there are possible sets of parameters
under which the minimum of the left part of $S(\rho)$ can be less than 2 and
the flow can have five different regimes of stability/instability. 
Nevertheless, under the set of parameters specified above we have up to three
regimes, where the third regime does not exist for $\rho''\geq \rho_{max}$
($A\leq 2D/T^2$).

Our numerical simulations show that $\rho_2\approx \rho''$, but $\rho'$ is
considerably greater than $\rho_1$, thus we expect that for
$\rho_1<\rho<\rho'$ both homogeneous and fluctuative states are stable.

In the fluctuative regime ($\rho_1<\rho<\rho_2$) the flow is
characterized by presence of humps (dense regions) moving backwards
or forwards. When the flow has stabilized the humps are equidistant and the evolution of
traffic in time and space resembles the spreading of a wave. The existence of a
fluctuative regime was predicted by other car-following (e.g. \cite{Sug} where
it was called "jammed flow") and continuous (e.g. \cite{LLK}, where it was
called "recurring humps state") models and measured experimentally \cite{KR1}. 

Simulations of our model
show that the fluctuative flow state is not unique.
Figs. 2(a-c) present the
cars velocities after the fluctuative flow regime has stabilized for three
different initial conditions. It can be seen
that the "wavelengths" of these states are different. Fig. 2(d) presents the
convergence of flux in these experiments to distinct values. Our
simulations also show the existence of solutions with other "wavelengths" and
flux values. Fig. 2(e) shows the fundamental diargams for three different
wavelengths. Consequently, depending on initial conditions different stable
fluctuative states emerge with different values of flux and
distances between neighboring humps. This indicates that for 
$\rho_1<\rho<\rho_2$  system (\ref{Initial}) has many stable periodic (in
$\Delta x_n$, $v_n$ variables) solutions, and hence in the $2N$-dimensional
space of variables $\Delta x_n$, $v_n$ there exist many attractive limit
cycles.

\begin{figure*}
\label{2}

\centerline{
\epsfig{file=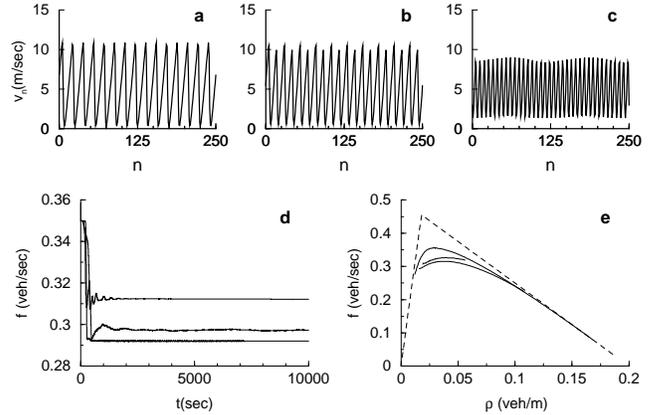, bb=170 35 590 655 , height=8.5cm, angle=270}
}
\vskip 0.3cm
\caption{ Three different stable states in the fluctuative regime, obtained
from different initial conditions. Global density $\rho=0.06(veh/m)$,
$A=3(m/s^2)$. (a-c) Cars velocities. (d) Convergence of
flux to different values in these three experiments. (e) Fundamental diagrams
for three different stable fluctuative states with wavelengths 20, 5 and 6.67
cars (top to bottom). A dashed line corresponds to the homogeneous solution.}

\end{figure*}

As follows from above for
$\rho_1<\rho<\rho'$ not only fluctuative flow solutions are stable, but also
the homogeneous flow solution. This bistability is the origin of
hysteresis in transitions between free and fluctuative flow regimes.
Such  bistability was observed experimentally \cite{KR1} and was found in 
other models \cite{Sug,PN,LLK}. Fig. 1(d) shows a hysteresis
loop in the density-flux plane. The upper curve is obtained by
increasing the density of cars adiabatically preserving the road
length $L$ \cite{insert}. It can be seen that up to the value of density
$\rho'$ the homogeneous flow is preserved. The lower
curve was obtained by adiabatically decreasing the density in the same manner.
While decreasing the density the flow remains fluctuative even for
$\rho<\rho'$. Fig.1(e) presents the hysteresis loop in the global density -
velocities fluctuations plane. 

Our results also illustrate the well-known phenomenon
\cite{KR1,KR2,Sug,LLK,Nagel} of jumps which the system performs in the
density-flux plane in the fluctuative flow regime when the density and the
flux are measured locally. In our numerical simulation (Fig. 1(f)) we started
from a value of density below $\rho'$, increased it gradually in the described
above manner up to a value greater than $\rho'$ and decreased it back.
These jumps may be explained by our finding of many stable states in the
fluctuative regime.

Our model also demonstrates the gradual spatial transition from the fluctuative
to free flow in the downstream direction which was measured by \cite{KR1}.  
The results of local measurements of density and flux at different distances
from an on-rump \cite{onrump} are shown in Fig.3. which is in good agreement
with Fig.3 of \cite{KR1}.


\begin{figure*}
\label{3}

\centerline{
\epsfig{file=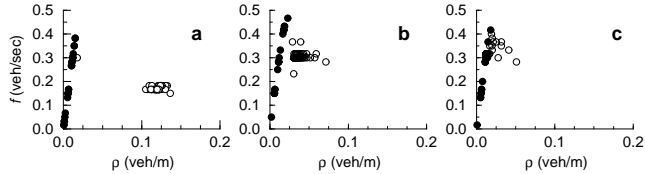, bb= 410 30 585 650, height=8.5cm, angle=270}
}
\vskip 0.2cm
\caption{Results of local measurements of flux and
density at different distances from the on-ramp. (a) 500 {\it m} upstream the
on-rump, (b) and (c) 250 and 1000 {\it m} downstream respectively}

\end{figure*}

In the congested flow regime the only stable solution is the homogeneous flow
solution. We have not found evidence of existence of bistability or hysteresis
in transitions between the fluctuative and congested flow regimes. Starting
from random initial conditions, we observe that initial fluctuations of the
velocity seem to decay according to a power law
\be
\sigma_v\sim\left\{
      \begin{array}{l}
       t^{-\beta},\quad t\ll t^* \\
       e^{-t/\tau},\quad t\gg t^*.
      \end{array}
      \right.
\en
where $t^*$ is the crossover time between the power law and exponential decay.
We find $t^* \sim L^z$ and $\tau \sim L^z$ with $z=2.0\pm 0.1$. These results
are qualitivly similar to that obtained by \cite{Kertesz} for a cellular
automata model \cite{NS}, but with different values of exponents. The result
$z \approx 2$ seems to be in agreement with random walk arguments of \cite{PN}.
For the parameters values $A=4, \quad \rho=0.15$ we get $\beta \approx 0.21\pm
0.04$ (Fig.4).

\begin{figure*}
\label{4}

\centerline{
\epsfig{file=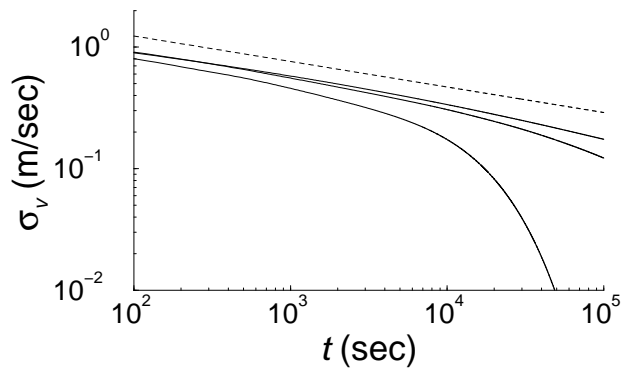, width=5cm, angle=270}
}
\caption{Decay of the velocities variance in time  in the congested flow
regime. Road length is 1, 4 and 16 km (bottom to top). The dashed line has a
slope -0.21}
\end{figure*}

In summary, we present a single lane car-following model which explains
important features of traffic observed experimentally. The model predicts the
existence of many stable periodic states in the fluctuative (synchronized)
flow regime.   

We wish to thank S. Schwarzer for usefull discussion.



\begin{references}
\bibitem{insert}
Every new car is placed in the middle between two consecutive
cars chosen randomly and gets the velocity of the car ahead. The safety time
gap of the new car and the car behind is halved and then gradually increased
linearly up to the initial value $T$.
\end{references}
\end{document}